\documentclass[aps,superscriptaddress,nofootinbib]{revtex4}

\usepackage{graphicx}
\usepackage{amsmath}
\usepackage{amssymb}

\newcommand{\ppara}{\ensuremath{p_\parallel}}

\newcommand{\Ppara}{\ensuremath{P_\parallel}}
\newcommand{\qpara}{\ensuremath{q_\parallel}}
\newcommand{\qperp}{\ensuremath{q_\perp}}

\newcommand{\gammapara}{\ensuremath{\gamma_\parallel}}
\begin{document}

\title{Is there a maximum magnetic field in QED?}

\author{Chung Ngoc Leung}
\email[E-mail: ]{leung@physics.udel.edu}
\affiliation{Department of Physics and Astronomy, University of Delaware, Newark, Delaware 19716, USA}

\author{Shang-Yung Wang}
\email[E-mail: ]{sywang@mail.tku.edu.tw}\thanks{corresponding author.}
\affiliation{Department of Physics, Tamkang University, Tamsui, Taipei 25137, Taiwan}

\date{\today}

\begin{abstract}
It was recently conjectured by Shabad and Usov that there exists in QED a maximum 
magnetic field of $10^{42}$~G, above which the magnetized vacuum becomes unstable. 
Using a nonperturbative analysis that consistently incorporates the effective electron 
mass and the screening effect in a strong magnetic field, we show that the conjectured 
phenomenon of positronium collapse never takes place. Thus, there does \emph{not} exist 
a maximum magnetic field in QED and the magnetized vacuum is stable for all values of 
the magnetic field.
\end{abstract}


\maketitle

A maximum value for the magnetic field in QED, $B_\mathrm{max}$, has recently been conjectured by Shabad and Usov~\cite{Shabad:2006ci}:
\begin{equation}
B_\mathrm{max}=\frac{m^2}{4e}\,\exp\bigg(\frac{\pi^{3/2}}{\sqrt{\alpha}}+2C_\mathrm{E}\bigg)\simeq 10^{42}~\mathrm{G},\label{Bmax}
\end{equation}
where $m$ and $e$ are respectively the electron mass and the absolute value of its charge in the absence of external fields, $\alpha=e^2/4\pi$ is the fine structure constant and $C_\mathrm{E}\simeq 0.577$ is Euler's constant. The maximum magnetic field of $10^{42}$~G, if correct, would rule out the existence of extremely strong magnetic fields of $10^{47}-10^{48}$~G in the vicinity of superconductive cosmic strings~\cite{Witten:1984eb}, and, most importantly, the exceeding of which would cause the restructuring of the strongly magnetized vacuum, thus calling for a revision of QED~\cite{Shabad:2005pn}.

In obtaining \eqref{Bmax} Shabad and Usov considered a positronium (i.e., an electron-positron bound state) placed in a strong magnetic field, and found that the magnetic field significantly enhances the Coulomb attraction between the constituent electron and positron. The Coulomb attraction becomes stronger and stronger until the electron and positron fall onto each other at the maximum magnetic field of $10^{42}$~G. To put it another way, when the magnetic field exceeds this maximum value, the rest energy of the positronium is fully compensated by the enormous binding energy such that the strongly magnetized positronium has a total energy less than that of the vacuum. This phenomenon is referred to by these authors as ``positronium collapse''~\cite{Shabad:2006ci} and could be a signal for possible vacuum instability~\cite{Shabad:2005pn}.

The aim of this Letter is twofold. First, we show that both the analysis leading to the conjecture as well as the conjecture itself are incorrect. This is because the phenomenon of magnetic catalysis of chiral symmetry breaking in QED in a strong magnetic field~\cite{Gusynin:1994re,Gusynin:1995gt,Leung:1996qy,Lee:1997zj,Gusynin:1998zq,Leung:2005yq,Wang:2007sn} is not properly taken into consideration\footnote{The same oversight also appears in two separate papers~\cite{Shabad:2007xu} of the same authors. See Ref.~\cite{Wang:2007jy} for a comment on the first paper of Ref.~\cite{Shabad:2007xu}}. Second, we present a nonperturbative analysis of the positronium that consistently incorporates the effective electron mass and the screening effect in a strong magnetic field, and show that the conjectured phenomenon of positronium collapse never takes place. As a consequence, there does not exist a maximum magnetic field in QED and the magnetized vacuum is stable for all values of the magnetic field.

While the existence of a maximum magnetic field in QED appears novel and is potentially of fundamental importance, it is in fact in contradiction with many of the well-established results in QED in a strong magnetic field~\cite{Gusynin:1994re,Gusynin:1995gt,Leung:1996qy,Lee:1997zj,Gusynin:1998zq,Leung:2005yq,Wang:2007sn}. In particular, magnetic catalysis of chiral symmetry breaking has long been known as a universal phenomenon. A strong magnetic field acts as a catalyst for chiral symmetry breaking, leading to the generation of a dynamical fermion mass even at the weakest attractive interaction between fermions. The hallmark of this phenomenon is the dimensional reduction from $(3+1)$ to $(1+1)$ in the dynamics of fermion pairing in a strong magnetic field when the lowest Landau level (LLL) plays the dominant role. The Nambu-Goldstone (NG) boson for spontaneously broken chiral symmetry is the massless fermion-antifermion bound state~\cite{Gusynin:1994re,Gusynin:1995gt,Leung:1996qy,Lee:1997zj}. Moreover, the phenomenon of magnetic catalysis is universal in that chiral symmetry is broken in arbitrarily strong magnetic fields and for any number of fermion flavors~\cite{Gusynin:1994re,Gusynin:1995gt,Leung:1996qy,Lee:1997zj,Gusynin:1998zq,Leung:2005yq}.

The realization of magnetic catalysis of chiral symmetry breaking in the chiral limit in QED (i.e., massless QED) has been studied extensively in the literature over the past decade~\cite{Gusynin:1995gt,Leung:1996qy,Lee:1997zj,Gusynin:1998zq,Leung:2005yq}. But until very recently, there has been no agreement on the correct calculation of the dynamical fermion mass generated through chiral symmetry breaking in QED in a strong magnetic field, and contradictory results have been found in the literature~\cite{Gusynin:1998zq,Gusynin:2002yi}. The resolution of the contradiction lies in the establishment of the gauge fixing independence of the dynamically generated fermion mass calculated using the Schwinger-Dyson equation approach~\cite{Leung:2005yq}. Furthermore, the universal nature of the phenomenon dictates that in realistic massive QED in a strong magnetic field, the electron will acquire a dynamical mass generated through the modification of the vacuum structure that is induced by the strong magnetic field. Indeed, a recent study shows that in massive QED the generation of a dynamical electron mass in a strong magnetic field is significantly enhanced by the perturbative electron mass that explicitly breaks chiral symmetry in the absence of a magnetic field~\cite{Wang:2007sn}. The effective electron mass in a strong magnetic field has been reliably determined. It is found~\cite{Wang:2007sn} that for asymptotically strong magnetic fields the effective electron mass, $m_\ast$, approaches from above its counterpart in massless QED, i.e., the dynamically generated fermion mass, $m_\mathrm{dyn}$.

The physics of magnetic catalysis of chiral symmetry breaking in realistic QED is essentially similar to that of chiral symmetry breaking in QCD~\cite{Klevansky:1992qe}, in which the quarks acquire constituent masses of dynamical origin that are brought about by the breaking of chiral symmetry. More importantly, the massive positronium is identified as the pseudo NG boson for explicit chiral symmetry breaking, pretty much in the same way the massive pion is identified in QCD. The universality of magnetic catalysis of chiral symmetry breaking, together with the Goldstone theorem, entails that the positronium has to appear in the spectrum of possible excitations in QED in arbitrarily strong magnetic fields. This in turn implies that there does not exist a maximum magnetic field in QED (be it massless or otherwise).

Taking into account the generation of an effective electron mass in QED in a strong magnetic field, we find the inequality in Ref.~\cite{Shabad:2006ci} that determines the condition for positronium collapse should be modified by (see (10) and (11) in the first paper, or (68) and (72) in the second paper,
of Ref.~\cite{Shabad:2006ci})
\begin{equation}
B\gg\frac{m_\ast^2(B)}{4 e}\,
\exp\bigg(\frac{\pi^{3/2}}{\sqrt{\alpha}}+2C_\mathrm{E}\bigg),\label{ineq1}
\end{equation}
where $m_\ast(B)$ is the effective electron mass in a strong magnetic field of magnitude $B$. Using the fact that $m_\ast(B)\approx m_\mathrm{dyn}(B)$ as $B\to\infty$~\cite{Wang:2007sn} and the result for $m_\mathrm{dyn}(B)$ in Ref.~\cite{Gusynin:1995gt} that was derived in the \emph{same} approximation as in obtaining \eqref{ineq1} [namely, the quenched rainbow (or ladder) approximation (RA)]:
\begin{equation}
m^\mathrm{(RA)}_\mathrm{dyn}(B)=C\,\sqrt{eB}\,
\exp\bigg(-\frac{\pi^{3/2}}{2\sqrt{2\alpha}}\bigg),\label{mdyn}
\end{equation}
where $C$ is a constant of order unity, we find upon substituting \eqref{mdyn} into \eqref{ineq1} that \eqref{ineq1} becomes (note that $B>0$)
\begin{equation}
\frac{C^2}{4}\,
\exp\bigg[\frac{\pi^{3/2}}{\sqrt{\alpha}}\bigg(1-\frac{1}{\sqrt{2}}\bigg)+2 C_\mathrm{E}\bigg]
\ll 1.\label{ineq3}
\end{equation}
Instead of obtaining an inequality that is trivially satisfied in the limit $B\to\infty$, we find that the inequality~\eqref{ineq3} cannot be satisfied for any reasonable values of $\alpha\ll 1$. Therefore, the conjectured phenomenon of positronium collapse never takes place and there does \emph{not} exist a maximum magnetic field in QED.

The absence of a maximum magnetic field in QED can also be understood by considering a positronium placed in a strong magnetic field. It is crucial to note that not only does the Coulomb attraction between the constituent electron and positron increase with increase of the magnetic field (as Shabad and Usov correctly pointed out), but so does the effective mass of the electron and positron (as Shabad and Usov failed to take into consideration). The net result of the two competing effects is that the increase in the effective mass is the dominant one, thanks to the wide \emph{separation of scales} in QED in a strong magnetic field
$m_\ast^{-1}(B)\approx m_\mathrm{dyn}^{-1}(B)\gg L_B\gg s_0^\mathrm{max}(B)$, where $L_B=1/\sqrt{eB}$ is the Larmor length and $s_0^\mathrm{max}(B)$ is the same as that defined in Ref.~\cite{Shabad:2006ci}, except with $m$ replaced by $m_\ast(B)$ in the definition. Therefore, the positronium is always separated from the vacuum by an energy gap and the conjectured positronium collapse never takes place. Furthermore, since $s_0^\mathrm{max}(B)$ is the length scale associated with the conjectured positronium collapse~\cite{Shabad:2006ci}, we emphasize that the absence of a maximum magnetic field confirms that, in consistence with the LLL dominance, the Larmor length serves as an ultraviolet regularization length scale in the problem in that the physics occurs at shorter distances is essentially irrelevant.

A powerful tool to study the electron-position bound states directly from QED is the Bethe-Salpeter (BS) equation truncated in a certain truncation scheme~\cite{Miransky}. For the problem at hand, however, in order to consistently incorporate the effective electron mass and the screening effect in a strong magnetic field, the electron and photon propagators that enter the BS equation must be obtained by solving the corresponding Schwinger-Dyson (SD) equations in the same truncation. This is tantamount to solving the truncated SD and BS equations simultaneously, one of the important points that has gone unnoticed in Ref.~\cite{Shabad:2006ci}.

Recently, it has been proved~\cite{Leung:2005yq,Wang:2007sn} that in QED (both massless and massive) in a strong magnetic field the bare vertex approximation (BVA), in which the vertex corrections are completely ignored, is a consistent truncation of the SD equations within the lowest Landau level approximation (LLLA). In particular, it can be shown that the truncated vacuum polarization is transverse and the dynamical fermion mass, obtained as the solution of the truncated fermion SD equation evaluated on the fermion mass shell, is manifestly gauge independent. Thus, for consistency with the results obtained in Refs.~\cite{Leung:2005yq,Wang:2007sn}, the BS equation for the positronium has to be truncated in the BVA within the LLLA.

\begin{figure}[t]
\begin{center}
\includegraphics[width=2.5truein,keepaspectratio=true,clip=true]{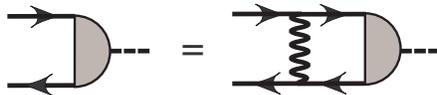}
\end{center}
\caption{BS equation in the BVA. The internal lines denote the full propagators, the vertex denotes the bare electron-photon vertex, and the filled half-circle represents the amputated BS amplitude for the positronium. External lines are amputated.}
\label{fig:BSBVA}
\end{figure}

We choose the constant external magnetic field of strength $B>0$ in the $x_3$-direction. The corresponding vector potential is given by $A^\mathrm{ext}_\mu=(0,0,Bx_1,0)$. Here, we will follow the conventions of Refs.~\cite{Leung:2005yq,Wang:2007sn}. In particular, the metric has the signature $g_{\mu\nu}=\mathrm{diag}(-1,1,1,1)$, the Dirac matrices satisfy $\{\gamma^\mu,\gamma^\nu\}=-2g^{\mu\nu}$ and $\gamma^5=i\gamma^0\gamma^1\gamma^2\gamma^3$. Note that the motion of the LLL electron and positron is restricted in directions perpendicular to the magnetic field, hence so is the motion of the bound state of the LLL electron and positron. We will henceforth refer to the latter as the LLL positronium. In the BVA within the LLLA, the BS equation for the LLL positronium is given by (see Fig.~\ref{fig:BSBVA})~\cite{Gusynin:1995gt,Lee:1997zj}
\begin{equation}
\chi(x,y;\Ppara)=-ie^2\int d^4x' d^4y'\gamma^\mu G(x,x')\chi(x',y';\Ppara)G(y',y)\gamma^\nu\mathcal{D}_{\mu\nu}(x,y),\label{chi1}
\end{equation}
where $\chi(x,y;\Ppara)$ is the amputated BS amplitude (wave function), i.e., the one-particle irreducible (1PI) electron-positron-positronium vertex with external legs amputated, and $\Ppara^\mu=(P^0,P^3)$ is the momentum of the LLL positronium. In \eqref{chi1}, $G(x,y)$ and $\mathcal{D}_{\mu\nu}(x,y)$ are respectively the \emph{full} propagators of the LLL electron and the photon in the external field $A^\mathrm{ext}_\mu$. The corresponding SD equations for $G(x,y)$ and $\mathcal{D}_{\mu\nu}(x,y)$ in the BVA within the LLLA have been studied in detail in Refs.~\cite{Leung:2005yq,Wang:2007sn}.

Utilizing the Ritus $E_p$ functions~\cite{Ritus} and following the analysis detailed in Ref.~\cite{Leung:2005yq}, we find \eqref{chi1} in momentum space (spanned by the $E_p$ functions) becomes
\begin{align}
\chi(\ppara;\Ppara)\,\Delta=&-ie^2\int_q\,\exp\biggl(-\frac{q_\perp^2}{2eB}\biggr)\,
\mathcal{D}_{\mu\nu}(q)\,\Delta\,\gammapara^\mu\,\Delta\,\frac{1}{\gammapara\cdot\ppara'+m_\ast}\,\chi(\ppara';\Ppara)\,
%\nonumber\\&\times
\frac{1}{\gammapara\cdot\ppara'+m_\ast}\,\Delta\,\gammapara^\nu\,\Delta,\label{chi2}
\end{align}
where $\ppara^\mu$ is the momentum of the LLL electron, $m_\ast$ is the effective electron mass in a strong magnetic field, $\ppara'=\ppara-\qpara$, $\qperp^2=q_1^2+q_2^2$ and $\int_q=\int d^4q/(2\pi)^4$. In \eqref{chi2}, $\Delta=(1+i\gamma^1\gamma^2)/2$ is the projection operator on the electron (positron) states with the spin polarized along (opposite to) the external magnetic field. Because the LLL electron and positron always have their spins polarized in opposite directions along the external magnetic field, the LLL positronium in its ground state is in fact a parapositronium, i.e., a pseudoscalar state. This, together with symmetry arguments, implies that the amputated BS amplitude $\chi(\ppara;\Ppara)$ takes the form
\begin{equation}
\chi(\ppara;\Ppara)=A(\ppara,\Ppara)\gamma^5,\label{chiform}
\end{equation}
where $A(\ppara,\Ppara)$ is a scalar function of $\ppara^2$ and $\Ppara^2$. Substituting \eqref{chiform} into \eqref{chi2}, we obtain
\begin{align}
A(\ppara,\Ppara)\gamma^5=&-ie^2\int_q\,\exp\biggl(-\frac{q_\perp^2}{2eB}\biggr)
\mathcal{D}_{\mu\nu}(q)A(\ppara',\Ppara)\gammapara^\mu\,\frac{1}{\gammapara\cdot\ppara'+m_\ast}\,\gamma^5\,
%\nonumber\\&\times
\frac{1}{\gammapara\cdot\ppara'+m_\ast}\,\gammapara^\nu,\label{A}
\end{align}
where use has been made of the properties $[\gammapara^\mu,\Delta]=[\gamma^5,\Delta]=0$, and the projection operator $\Delta$ that multiplies both sides of \eqref{A} has been dropped. When evaluated on the respective particle mass shells, i.e., $\ppara^2=-m_\ast^2$ and $\Ppara^2=-M^2$, and supplemented with the effective electron mass $m_\ast$ that is obtained as the solution of the on-shell SD equations~\cite{Leung:2005yq,Wang:2007sn}, the BS equation \eqref{A} can be used to determine the mass of the LLL positronium, $M$.

Before proceeding further, let us discuss the gauge (in)dependence of the on-shell BS equation. This is an issue of fundamental importance that was not addressed in Ref.~\cite{Shabad:2006ci}. It is noted that the function $A(\ppara,\Ppara)$ in \eqref{A} depends implicitly on the gauge fixing parameter through the full photon propagator $\mathcal{D}_{\mu\nu}(q)$. The latter in covariant gauges is given by~\cite{Gusynin:1998zq,Leung:2005yq,Wang:2007sn}
\begin{equation}
\mathcal{D}^{\mu\nu}(q)=\frac{1}{q^2+\Pi(\qpara^2,\qperp^2)}
\biggl(g_\parallel^{\mu\nu}-\frac{q^\mu_\parallel
q^\nu_\parallel}{\qpara^2}\biggr)+\frac{g_\perp^{\mu\nu}}{q^2}+\frac{q^\mu_\parallel q^\nu_\parallel}{q^2\qpara^2}
+(\xi-1)\frac{1}{q^2}\frac{q^\mu q^\nu}{q^2},\label{D}
\end{equation}
where $\Pi(\qpara^2,\qperp^2)$ is the polarization function (see Refs.~\cite{Gusynin:1998zq,Leung:2005yq,Wang:2007sn} for explicit expression) and $\xi$ is the gauge fixing parameter with $\xi=1$ being the Feynman gauge. We emphasize that because of this gauge dependence, great care must be taken in analyzing the on-shell BS equation and the LLL positronium mass extracted therefrom.

\begin{figure}[t]
\begin{center}
\includegraphics[width=3.0truein,keepaspectratio=true,clip=true]{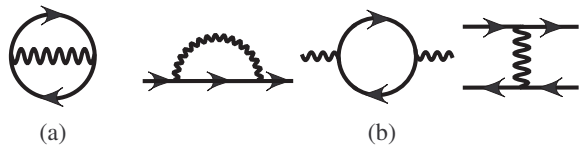}
\end{center}
\caption{(a) Contribution to the 2PI effective action at two-loop order in the loop expansion. (b) The electron self-energy, vacuum polarization and electron-positron interaction kernel generated thereby. External lines are amputated.}
\label{fig:2PI}
\end{figure}

A detailed analysis based on the Ward-Takahashi identity in the BVA within the LLLA reveals that contrary to the gauge independence of the on-shell SD equations~\cite{Leung:2005yq,Wang:2007sn}, the on-shell BS equation truncated in the BVA is inevitably gauge dependent. At this point it might appear that there is no way to reliably determine the properties of the positronium in the BVA within the LLLA. We now argue that this is not the case. In particular, in the BVA the on-shell BS equation has a controlled gauge dependence, thanks to a direct correspondence~\cite{Leung:2005yq} between the SD equations truncated in the BVA and the 2PI effective action~\cite{Cornwall:1974vz} truncated at the lowest nontrivial (two-loop) order in the loop expansion. Let the functional $\Gamma_2[G,\mathcal{D}]$ denotes the sum of all 2PI skeleton vacuum diagrams with \emph{bare} vertex and \emph{full} LLL electron and photon propagators. The contribution to $\Gamma_2[G,\mathcal{D}]$ at two-loop order is depicted diagrammatically in Fig.~\ref{fig:2PI}(a). The key point of the argument is to note that the direct correspondence can be generalized to include the BS equation truncated in the BVA. This is because, as shown in Fig.~\ref{fig:2PI}(b), the corresponding electron self-energy, vacuum polarization and electron-positron interaction kernel that enter the SD and BS equations in the BVA are the same as those generated by $\Gamma_2[G,\mathcal{D}]$. The argument is completed with the fact~\cite{Arrizabalaga:2002hn} that the truncated 2PI effective action evaluated at its stationary point has a controlled gauge dependence, i.e., the explicit gauge dependent terms always appear at higher order. Moreover, since the transverse components in $\mathcal{D}^{\mu\nu}(q)$ decouple and the gauge dependent contribution in \eqref{A} arises from the longitudinal components in $\mathcal{D}^{\mu\nu}(q)$ proportional to $q^\mu q^\nu /q^2$, we conclude that in the BVA and at the order of truncation only the first term in $\mathcal{D}^{\mu\nu}(q)$ proportional to $g_\parallel^{\mu\nu}$ contributes to the on-shell BS equation.

We are now in position to prove the nonexistence of a maximum magnetic field in QED. For this purpose, we can simply consider the on-shell BS equation in an asymptotically strong magnetic field. For $B\gg B_0\equiv m^2/e$, the explicit breaking of chiral symmetry associated with the perturbative electron mass $m$ can be neglected. As per the fact that $m_\ast\approx m_\mathrm{dyn}$ as $B\to\infty$~\cite{Wang:2007sn} and the pseudo NG boson nature of the positronium that $M\to 0$ as $B\to\infty$, the BS equation \eqref{A} is evaluated on the mass shells $\ppara^2=-m_\mathrm{dyn}^2$ and $\Ppara^2=0$. Moreover, because of the strong screening effect in a strong magnetic field~\cite{Gusynin:1998zq,Leung:2005yq,Wang:2007sn}, the integral in \eqref{A} is dominated by contributions from the region with $m_\mathrm{dyn}^2\lesssim |\qpara^2|\ll eB$. Thus, $A(\ppara',\Ppara)$ in the integrand can be approximated by $A(\ppara,\Ppara)$. After some algebra, we find that in the limit $B\to\infty$ the on-shell BS equation reduces to
\begin{equation}
1=-2ie^2\int_q\,\frac{\exp(-q_\perp^2/2eB)}{q^2+\Pi(\qpara^2,\qperp^2)}
\frac{1}{(p-q)_\parallel^2+m_\mathrm{dyn}^2}\bigg|_{\ppara^2=-m_\mathrm{dyn}^2}.\label{gap}
\end{equation}
The proof is completed by noting that \eqref{gap} is the same as the on-shell SD equation obtained
in the BVA within the LLLA~\cite{Leung:2005yq} that reliably determines the dynamically generated fermion mass, $m_\mathrm{dyn}$, in massless QED (see (4.20) in the first paper of Ref.~\cite{Leung:2005yq}). This also serves to justify \emph{a posteriori} the controlled gauge dependence of the on-shell BS equation.

In conclusion, the positronium is unambiguously identified as the (pseudo) Nambu-Goldstone boson for spontaneous (explicit) chiral symmetry breaking in massless (massive) QED in a strong magnetic field. It is shown that the phenomenon of positronium collapse conjectured by Shabad and Usov in Ref.~\cite{Shabad:2006ci} never takes place. Consequently, there does not exist a maximum magnetic field in QED and the magnetized vacuum is stable for all values of the magnetic field.

\section*{Acknowledgements}

S.-Y.W.\ thanks A.\ E.\ Shabad for email correspondences during the initial stages of this work. This research was supported in part by the National Science Council of Taiwan under grant 96-2112-M-032-005-MY3.

\end{document}